\documentclass[rnaas]{aastex62}

\begin{document}

\title{The Observed Rate of Binary Black Hole Mergers can be Entirely Explained by Globular Clusters}

%% Note that the corresponding author command and emails has to come
%% before everything else. Also place all the emails in the \email
%% command instead of using multiple \email calls.
\correspondingauthor{Carl L.~Rodriguez}
\email{carlrodriguez@cmu.edu}

\author[0000-0003-4175-8881]{Carl L. Rodriguez}
\affil{McWilliams Center for Cosmology and Department of Physics, Carnegie Mellon University, Pittsburgh, PA 15213, USA}

\author[0000-0002-4086-3180]{Kyle Kremer}
\affiliation{TAPIR, California Institute of Technology, Pasadena, CA 91125, USA}
\affiliation{The Observatories of the Carnegie Institution for Science, Pasadena, CA 91101, USA}

\author[0000-0002-3680-2684]{Sourav Chatterjee}
\affil{Tata Institute of Fundamental Research, Homi Bhabha Road, Navy Nagar, Colaba, Mumbai 400005, India}

\author[0000-0002-7330-027X]{Giacomo Fragione}
\affil{Center for Interdisciplinary Exploration \& Research in Astrophysics (CIERA) and Department of Physics \& Astronomy, Northwestern University, Evanston, IL 60208, USA}

\author{Abraham Loeb}
\affil{Astronomy Department, Harvard University, 60 Garden Street, Cambridge, MA 02138}

\author[0000-0002-7132-418X]{Frederic A. Rasio}
\affil{Center for Interdisciplinary Exploration \& Research in Astrophysics (CIERA) and Department of Physics \& Astronomy, Northwestern University, Evanston, IL 60208, USA}

\author[0000-0002-9660-9085]{Newlin C. Weatherford}
\affil{Center for Interdisciplinary Exploration \& Research in Astrophysics (CIERA) and Department of Physics \& Astronomy, Northwestern University, Evanston, IL 60208, USA}

\author[0000-0001-9582-881X]{Claire S. Ye}
\affil{Center for Interdisciplinary Exploration \& Research in Astrophysics (CIERA) and Department of Physics \& Astronomy, Northwestern University, Evanston, IL 60208, USA}

\begin{abstract}
Since the first signal in 2015, the gravitational-wave detections of merging binary black holes (BBHs) by the LIGO and Virgo collaborations (LVC) have 
completely transformed our understanding of the lives and deaths of compact object binaries, and have motivated an enormous amount of theoretical work on the 
astrophysical origin of these objects.  We show that the phenomenological fit to the redshift-dependent merger rate of BBHs from \cite{poppaper} is consistent 
with a purely dynamical origin for these objects, and that the current merger rate of BBHs from the LVC could be explained entirely with globular clusters 
alone.  While this does not prove that globular clusters are the dominant formation channel, we emphasize that many formation scenarios could contribute a significant fraction of the current LVC rate, and that any analysis that assumes a single (or dominant) mechanism for producing BBH mergers is implicitly using a specious astrophysical prior.
\end{abstract}


%% Start the main body of the article. If no sections in the 
%% research note leave the \section call blank to make the title.
\section{}

Over the past 5 years, the observed rate of BBH mergers in the local universe has been significantly constrained.  After the detections of GW150914, GW151012, 
and GW151226 in the first observing run, initial estimates put the volumetric merger rate between 9 to $240~\rm{Gpc}^{-3}\rm{yr}^{-1}$ at 90\% confidence at 
$z=0$ \citep{Abbott2016e}.  Merger rates $\gtrsim 100~\rm{Gpc}^{-3}\rm{yr}^{-1}$ were largely understood to be explicable by isolated binary evolution 
\citep[e.g.,][]{Belczynski2016a}, with dynamical assembly \citep[e.g.,][]{Rodriguez2016a} only contributing a small fraction of the population.  But as more BBH 
mergers have been detected, the measured local merger rate has decreased by an order of magnitude, while sufficient numbers of higher-redshift mergers allow the 
slope of the rate to be observed.  By fitting the observed detections to a phenomenological model of the form $R(z)=R_0(1+z)^\kappa $, an analysis of the latest 
GW transient catalog \cite[GWTC-2,][]{poppaper} suggests an increasing merger rate with a local value of $19^{+16}_{-9}~\rm{Gpc}^{-3}\rm{yr}^{-1}$ at $z=0$.

%% An example figure call using \includegraphics
\begin{figure*}[t!]
\begin{center}
\includegraphics[scale=0.85,trim=48 0 10 0,clip]{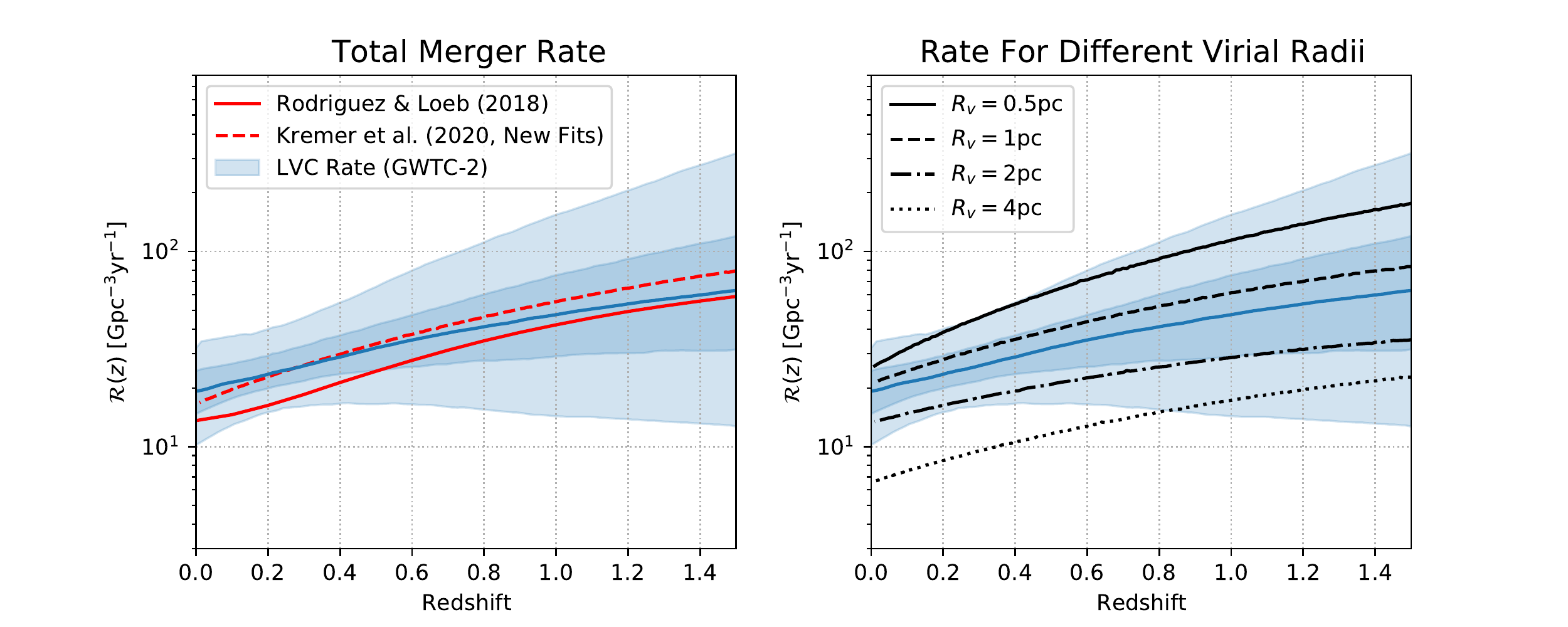}
\caption{The merger rate of BBHs from GCs compared to the latest LVC results.  The \textbf{left} panel shows the predictions from \cite{Rodriguez2018b} and updated predictions from \cite{Kremer2020} using the same cosmological model.  The \textbf{right} panel shows the fits to the models from \cite{Kremer2020} broken down according to cluster virial radius, assuming that all clusters are born with initial radii of 0.5, 1, 2, or 4 pc.  The total prediction on the left is the average of these four values.  In blue, we show the median, 50\%, and 90\% intervals of the phenomenological merger rate fit from GWTC-2 \citep{poppaper}.}
\end{center}
\end{figure*}

There have been many theoretical models of the merger rate of BBHs from GCs, both before and after the first detection of GWs 
\cite[e.g.][]{PortegiesZwart2000,Rodriguez2015a} %,OLeary2006,Moody2009,Downing2010,Tanikawa2013,Bae2014,Rodriguez2015a,Rodriguez2016,Askar2016,Rodriguez2018,2019ApJ...873..100C,Kremer2020,2020PhRvD.102l3016A}, 
with the majority predicting a volumetric merger rate at $z=0$ of $\sim10~\rm{Gpc}^{-3}\rm{yr}^{-1}$.  Using the Cluster Monte Carlo code (CMC), 
several groups have studied the BBH merger rate from GCs and the unique properties of their sources.  The analysis of the BBH merger rate from GCs initially predicted a merger rate anywhere from 2 to $20~\rm{Gpc}^{-3}\rm{yr}^{-1}$ at $z=0$, based on the luminosity function and comoving spatial density of observed GCs in the local universe \citep{Rodriguez2016a}.   Of course, this initial analysis ignored the contributions of GCs that were disrupted before the present day, which can significantly increase the contribution to the BBH merger rate \cite[e.g.,][]{Fragione2018}.  In \cite{Rodriguez2018}, we combined a cosmological model for GC formation \citep{El-Badry2018} with star-by-star CMC models of GCs to estimate the BBH merger rate as a function of cosmological redshift.

In the left panel of Figure 1, we show the predictions from \cite{Rodriguez2018} and updated predictions using the same cosmological model and fitting 
procedure\footnote{For the updated model we have added an additional correction accounting for the formation of central-massive BHs in GCs, from \cite{2019MNRAS.486.5008A}; 
see \cite{2019ApJ...881...75K} for details.  This decreases the BBH merger rate from the densest GCs by $\sim 10\%$} with newer GC models (covering a wider 
range of parameters) from \cite{Kremer2020}.  When compared to the allowed range and cosmological evolution of the BBH merger rate from GWTC-2 \citep{poppaper}, it is obvious that \emph{the entire phenomenological BBH merger rate can potentially be explained by GCs alone}.  At $z=0$ the original fits predict a merger rate of $15~\rm{Gpc}^{-3}\rm{yr}^{-1}$, while the newer fits from \cite{Kremer2020} predict a merger rate of $17~\rm{Gpc}^{-3}\rm{yr}^{-1}$.  In both cases, the merger rate increases as a function of redshift.  Comparing the ratio of the rate at $z=1$ to $z=0$, we find that $R(1)/R(0) = 3.1$ for the original fits from \cite{Rodriguez2018}, and $R(1)/R(0) = 3.2$ for the \cite{Kremer2020} models.  Both ratios are consistent with the phenomenological value of $R(1)/R(0) = 2.5^{+7.8}_{-1.9}$ from \cite{poppaper}.  

The primary improvement in the models of \cite{Kremer2020} is the extent of the grid, with the newer grid containing clusters with higher initial masses, more compact initial virial radii (as small as 0.5 pc), and a wider range of metallicities.  All together, they capture nearly the complete spectrum of present-day dense star clusters in the Milky Way.  We have assumed in the left panel of Figure 1 that clusters of different initial radii contribute equally to the total rate.  To make this more explicit, we show what the merger rate would look like from GCs assuming all cluster were born with virial radii of 0.5, 1, 2, or 4 pc in the right panel of Figure 1.  The four values clearly span the 90\% region of allowed merger rates from \cite{poppaper}.  As pointed out in \cite{Kremer2020}, initial concentrations of clusters directly control the slope of the merger rate, with the contributions from clusters with 0.5, 1, 2, and 4 pc having ratios of $R(1)/R(0) = 4.2$, 2.8, 2.1, and 2.6, respectively.  Given that many young clusters in the local universe are observed to have initial effective radii between 1 and 2 pc \cite[e.g.][]{scheepmaker2007}, this suggests remarkably good agreement with current LVC findings.

Of course, there are many such dynamical scenarios for forming merging BBHs %, including mergers in smaller, open clusters \cite[e.g.,][]{Ziosi2014,Banerjee2017,2019MNRAS.487.2947D}, triples in star clusters \citep[e.g.,][]{Wen2003,2016ApJ...816...65A,Hoang2018,2020ApJ...903...67M}, or from stellar triples and quadruples \citep[e.g.,][]{2017ApJ...841...77A,ll18,2018ApJ...863....7R,fragk2019}, migration and captures in AGN disks \cite[e.g.,][]{Stone2016,Bartos2016}, dynamical encounters in galactic fields \cite[e.g.,][]{2019ApJ...887L..36M}
, many of which are consistent with the 90\% uncertainties from the LVC's phenomenological model.  But while the measured merger rates can now be explained by a single formation channel alone, there is no reason to believe that is the case.  Several studies of the current LVC BBH catalog have suggested that multiple formation scenarios likely operate in producing BBH mergers \cite[e.g.,][]{2020arXiv201103564W,2020arXiv201110057Z} with the latter suggesting that no single channel likely contributes more than 70\% to the total population. This note is meant to emphasize this point: given that GCs alone can naturally explain the most up-to-date BBH merger rate from the LVC, it is no longer reasonable to assume that dynamical processes constitute a subdominant fraction of the full merger rate.

\acknowledgments

% \software{\texttt{numpy} \citep{walt_numpy_2011}, \texttt{scipy} \citep{2020SciPy}, \texttt{matplotlib} \citep{hunter_matplotlib:_2007}}

%\bibliographystyle{aasjournal}
%\bibliography{biblio}

\end{document}